\documentclass[twocolumn,citeautoscript,showpacs,preprintnumbers,amsmath,amssymb,prb,floatfix,footinbib]{revtex4}
\usepackage{graphicx}
\usepackage{dcolumn}
\usepackage{bm}
\usepackage{times}
\begin{document}
\title{Suppression of antiferromagnetic order and orthorhombic distortion in superconducting Ba(Fe$_{0.961}$Rh$_{0.039}$)$_2$As$_2$}
\author{A.~Kreyssig, M.\,G.~Kim, S.~Nandi, D.\,K.~Pratt, W.~Tian, J.\,L.~Zarestky, N.~Ni, A.~Thaler, S.\,L.~Bud'ko, P.\,C.~Canfield, R.\,J.~McQueeney, and A.\,I.~Goldman}
\affiliation{\\Ames Laboratory, U.\,S. DOE and Department of Physics and Astronomy\\
Iowa State University, Ames, IA 50011, USA}

\date{\today}

\begin{abstract}
Neutron diffraction and high-resolution x-ray diffraction studies 
find that, similar to the closely related underdoped 
Ba(Fe$_{1-x}$Co$_x$)$_2$As$_2$ superconducting compounds, 
Ba(Fe$_{0.961}$Rh$_{0.039}$)$_2$As$_2$ shows strong evidence of 
competition and coexistence between superconductivity and 
antiferromagnetic order below the superconducting transition, 
$T_c$~= 14\,K.  The transition temperatures for both the magnetic 
order and orthorhombic distortion are in excellent agreement with 
those inferred from resistivity measurements, and both order 
parameters manifest a distinct decrease in magnitude below 
$T_c$.  These data suggest that the strong interaction between 
magnetism and superconductivity is a general feature of 
electron-doped Ba(Fe$_{1-x}T\!M_x$)$_2$As$_2$ superconductors 
($T\!M$~= Transition Metal).
\end{abstract}


\maketitle 

Recent systematic neutron and x-ray diffraction studies of the 
underdoped Ba(Fe$_{1-x}$Co$_x$)$_2$As$_2$ superconductors have 
revealed several startling results regarding the interactions 
among structure, magnetism and superconductivity.  The undoped 
$A\!E$Fe$_2$As$_2$ parent compounds ($A\!E$~= Ba, Sr, or Ca) 
manifest simultaneous transitions from a high-temperature 
paramagnetic tetragonal phase to a low-temperature orthorhombic 
antiferromagnetic (AFM) 
structure\cite{Huang08,Jesche08,Goldman08,Ni08}.  Upon doping 
with Co for Fe in Ba(Fe$_{1-x}$Co$_x$)$_2$As$_2$\cite{Sefat08}, 
however, both the structural ($T_S$) and antiferromagnetic order 
($T_N$) transitions are suppressed to lower temperatures and 
split, with $T_S$ slightly higher than 
$T_N$\cite{Ni08b,Chu09,Lester09}.  Several neutron and x-ray 
studies have clearly established that both the magnetic ordering 
and orthorhombic distortion are distinctly sensitive to 
superconductivity throughout the Ba(Fe$_{1-x}$Co$_x$)$_2$As$_2$ 
series\cite{Lester09,Pratt09,Christianson09,Fernandes10,Nandi10}.  
At a given Co composition, as the sample temperature is reduced 
below $T_c$, there is a clear suppression of the magnetic order 
parameter and, in fact, reentrance into the paramagnetic phase is 
observed for a Co-doping concentration of 
$x$~=~0.059\cite{Fernandes10}.  Similarly, the magnitude of the 
orthorhombic lattice distortion, as evidenced by high-resolution 
x-ray diffraction measurements of peak splitting in 
Ba(Fe$_{1-x}$Co$_x$)$_2$As$_2$, decreases below $T_c$ and 
reentrance into the tetragonal structure was observed at 
Co-doping concentrations of $x$~=~0.063\cite{Nandi10}.  This 
striking behavior for Ba(Fe$_{1-x}$Co$_x$)$_2$As$_2$ is 
unprecedented and has been related to the strong coupling between 
superconductivity and magnetism\cite{Fernandes10} as well as an 
unusual magnetoelastic coupling in the form of emergent nematic 
order in the iron 
arsenides\cite{Nandi10,Fang08,Xu08,Fernandes10b}.  

This unusual behavior has not, until now, been reported for other 
doping species in the Ba(Fe$_{1-x}T\!M_x$)$_2$As$_2$ family of 
compounds, but is widely suspected.  Studies of bulk 
thermodynamic and transport properties of 
Ba(Fe$_{1-x}T\!M_x$)$_2$As$_2$, for $T\!M$~= Co, Ni, Cu, Rh, or 
Pd, have recently been reviewed in some detail\cite{Canfield10} 
and noted that superconductivity appears when two conditions are 
satisfied: (i) the structural and magnetic phase transitions must 
be suppressed to sufficiently low temperatures while (ii) the 
number of additional electrons contributed by the doping falls 
within a specific window.  These points are supported by, for 
example, the absence of superconductivity for 
Cu-doping\cite{Canfield09}.  Nevertheless, the behavior of the 
structural and magnetic transitions for all doping species, in 
the underdoped regime, is quite similar to what has been found 
for the case of Co-doping; both transitions are suppressed in 
temperature and split.  It is important to establish, if the 
behavior of the magnetic and structural order parameters in the 
superconducting state also carries over to the other doping 
species since differences, such as steric effects or disorder, may 
also play a role.  

Here we report on neutron diffraction measurements and 
high-resolution x-ray diffraction measurements on 
Ba(Fe$_{0.961}$Rh$_{0.039}$)$_2$As$_2$ that confirm the same 
unusual behavior of the magnetic and structural order parameters 
below $T_c$.  This composition was chosen for a direct comparison 
with previous results on 
Ba(Fe$_{0.953}$Co$_{0.047}$)$_2$As$_2$\cite{Pratt09}, i.\,e. the 
structural, magnetic and superconducting transition temperatures 
for Ba(Fe$_{0.953}$Co$_{0.047}$)$_2$As$_2$ are $T_S$~= 60\,K, 
$T_N$~= 47\,K and $T_c$~= 17\,K respectively, and are close to 
the corresponding transition temperatures, $T_S$~= 59\,K, $T_N$~= 
54\,K and $T_c$~= 14\,K, for 
Ba(Fe$_{0.961}$Rh$_{0.039}$)$_2$As$_2$. 

\begin{figure}
\includegraphics[width=1.0\linewidth]{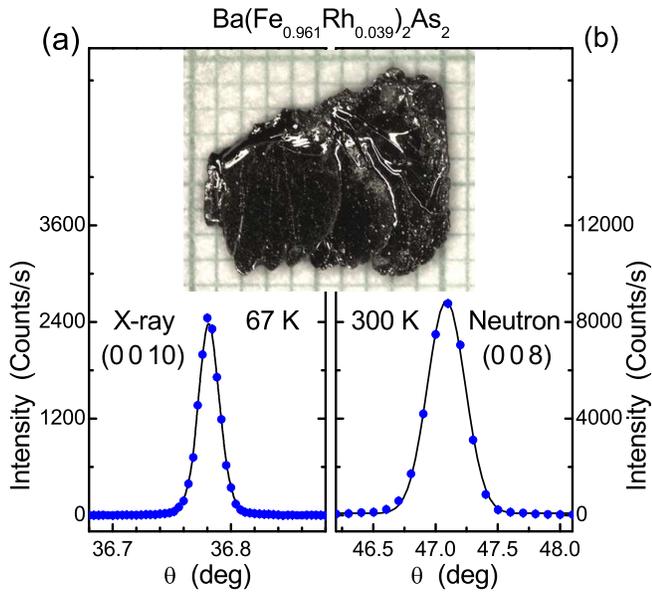}
\caption{\label{fig:fig1}(a) X-ray and (b) neutron rocking curves 
of the Ba(Fe$_{0.961}$Rh$_{0.039}$)$_2$As$_2$ sample investigated 
in this study.  The picture shows the single crystal on a 
millimeter grid.}
\end{figure}

Single crystals of Ba(Fe$_{0.961}$Rh$_{0.039}$)$_2$As$_2$ were 
grown out of a FeAs self-flux using conventional high-temperature 
solution growth\cite{Ni09}. The compositions were measured at 18 
positions on several samples of the used batch by wavelength 
dispersive spectroscopy to determine the Rh-doping composition as 
$x$~=~0.039$\pm$0.002.  The samples emerge from the flux as 
plates with the tetragonal \textbf{c} axis perpendicular to the 
plate.  The particular sample used for the neutron and x-ray 
measurements (mass of 260\,mg) is shown as inset to Fig.~1, and 
exhibited a sharp sample mosaic in both x-ray [Fig.~1(a)] and 
neutron [Fig.~1(b)] rocking scans, demonstrating excellent sample 
quality. Magnetization and temperature-dependent AC electrical 
resistance data ($f$~= 16\,Hz, $I$~= 3\,mA) were collected in a 
Quantum Design Magnetic Properties Measurement System using a 
Linear Research LR700 resistance bridge for the latter. 
Electrical contact was made to the sample using Epotek H20E 
silver epoxy to attach Pt wires in a four-probe configuration. As 
shown in Figs.~2(a) and (b), both the magnetization and 
resistivity measurements show the onset of superconductivity in 
the sample at approximately 14\,K, whereas the structural and 
magnetic transitions at higher temperatures are only weakly in 
evidence, but can be discerned in the derivative of the 
resistivity. 

\begin{figure}
\includegraphics[width=1.0\linewidth]{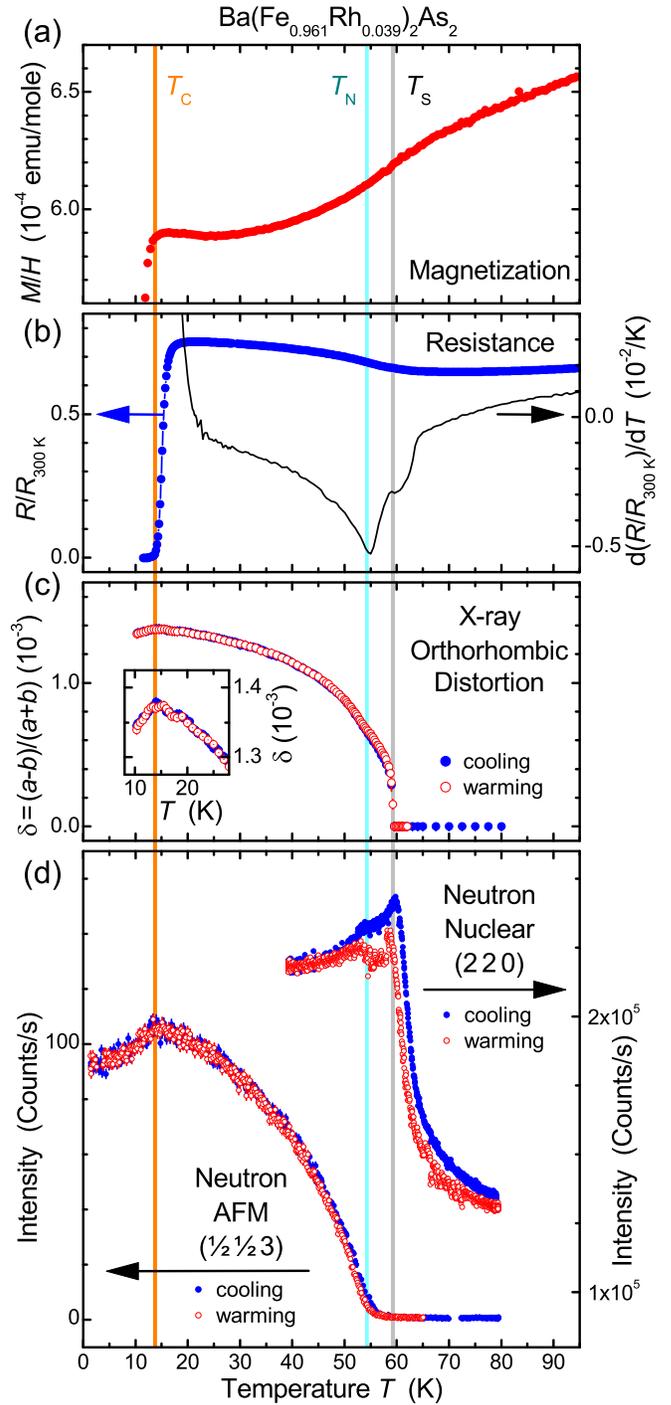}
\caption{\label{fig:fig2}(a) Measured magnetization of 
Ba(Fe$_{0.961}$Rh$_{0.039}$)$_2$As$_2$ showing the 
superconducting transition at $T_c$~= 14\,K. (b) Resistance 
(points) and its derivative (line) showing the signature of 
$T_S$, $T_N$ and $T_c$. (c) Orthorhombic distortion measured as a 
function of temperature on cooling and warming.  The inset shows 
an expanded region close to $T_c$ to emphasize the decrease of 
the orthorhombicity in the superconducting state. (d) Intensity 
of the nuclear (2\,2\,0) reflection as a function of temperature 
close to $T_S$ on cooling and warming, and the intensity of the 
(1/2\,1/2\,3) magnetic reflection on cooling and warming.}
\end{figure}

To elucidate the nature of the structural transition and its 
behavior below $T_c$, high-resolution single-crystal x-ray 
diffraction measurements were performed on a four-circle 
diffractometer using Cu-$K_{\alpha1}$ radiation from a rotating 
anode x-ray source, selected by a germanium (1\,1\,1) 
monochromator. For these measurements, the plate-like single 
crystal was attached to a flat copper sample holder on the cold 
finger of a closed-cycle displex refrigerator. The mosaicity of 
the Ba(Fe$_{0.961}$Rh$_{0.039}$)$_2$As$_2$ single crystal was 
less than 0.02\,deg full-width-at-half-maximum as measured by the 
rocking curve of the (0\,0\,10) reflection at room temperature 
[Fig.~1(a)]. The diffraction data were obtained as a function of 
temperature between room temperature and 10\,K, the base 
temperature of the refrigerator.  Fig.~3(a) shows a subset of 
($\xi\,\xi$\,0) scans through the (1\,1\,10) reflection as the 
sample was cooled through $T_S$~= 59\,K. The splitting of the 
peak below $T_S$ is consistent with the structural transition, 
from space group $I\,4/m\,m\,m$ to $F\,m\,m\,m$, with a 
distortion along the [1\,1\,0] direction. As the sample is cooled 
further, the orthorhombic splitting increases until $T_c$~= 
14\,K. Lowering the temperature below $T_c$ results in a decrease 
in the orthorhombic distortion consistent with what was 
previously observed for 
Ba(Fe$_{1-x}$Co$_x$)$_2$As$_2$\cite{Nandi10}. The extrapolated 
value (1.4$\pm$0.1)$\cdot$10$^{-3}$ for $T$~= 0\,K is also the 
same, within experimental error, as observed in 
Ba(Fe$_{0.953}$Co$_{0.047}$)$_2$As$_2$\cite{Nandi10}. 

Neutron diffraction measurements were performed on the HB1A 
diffractometer at the High Flux Isotope Reactor at Oak Ridge 
National Laboratory using a sample with a mass of 260\,mg and a 
resolution limited crystal mosaic ($<$\,0.3\,deg 
full-width-at-half-maximum). The experimental configuration was 
48'-40'-40'-136' with fixed incident neutron energy of 14.7\,meV, 
and two pyrolytic graphite filters for the elimination of higher 
harmonics in the incident beam.  The sample was aligned in the 
($H\,H\,L$) scattering plane and mounted in a closed-cycle 
refrigerator. 

\begin{figure}
\includegraphics[width=1.0\linewidth]{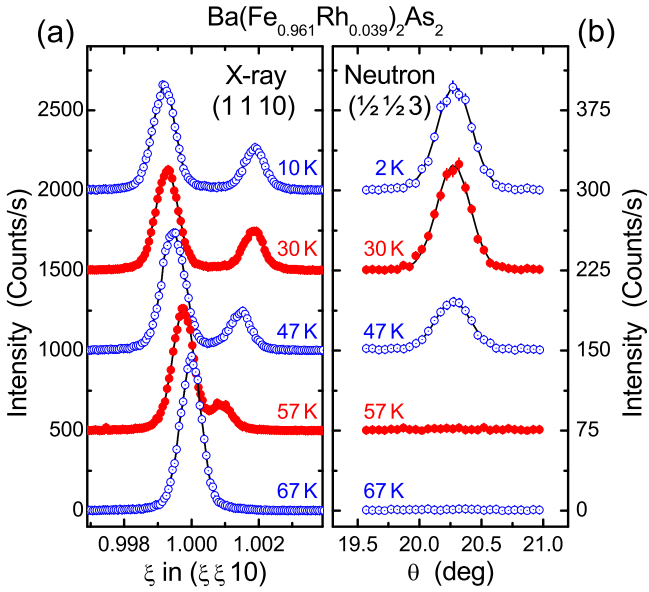}
\caption{\label{fig:fig3}Transverse scans through (a) the 
(1\,1\,10) charge reflection and (b) the (1/2\,1/2\,3) magnetic 
reflection demonstrating the transition between the tetragonal 
and orthorhombic structures and the onset of AFM order, 
respectively.}
\end{figure}

Even though the instrumental resolution of the neutron 
spectrometer was insufficient to resolve the splitting of nuclear 
peaks at the tetragonal-to-orthorhombic transition, the 
structural distortion can be discerned through the behavior of 
the nuclear peaks in the vicinity of $T_S$.  Figure~2(d) shows 
the evolution of the integrated intensity of the (2\,2\,0) 
nuclear reflection with temperature. We find that as the 
temperature decreases below approximately 65\,K, the intensity of 
the (2\,2\,0) reflection increases sharply and peaks at 59\,K. 
The peak is observed at the same transition temperature, $T_S$, 
determined in the high-resolution x-ray study. The increase in 
intensity of the (2\,2\,0) reflection arises from the extinction 
release that occurs due to stress, strain and domain formation 
related to the orthorhombic distortion. This effect is strongest 
at the transition because the fluctuations are strongest, leading 
to a maximum in local stress and strain which decreases above and 
below the transition. The observation of measurable extinction 
release at temperatures well above $T_S$ indicates the presence 
of significant structural fluctuations related to the 
orthorhombic distortion. Indeed, this effect sets in at roughly 
the same temperature where deviations of transport properties 
[see Fig.~2(b)] from their high-temperature behavior are first 
observed, pointing to significant electron scattering by the 
structural fluctuations.  Below the structural transition, the 
orthorhombic lattice distortion leads to twin domains that grow 
in size with decreasing temperature. The impact of these twin 
domains on the nuclear (2\,2\,0) reflection is demonstrated by 
the occurrence of an additional distinct feature in the 
temperature dependence of the intensity at $T_N$~= 54\,K.  At 
this temperature, due to strong magnetoelastic 
coupling\cite{Nandi10,Fernandes10b}, the domain structure changes 
again at the onset of magnetic order.  The high quality of the 
sample is, again, evidenced by the relatively strong extinction 
release at the structural transition and the fact that even the 
mild change in extinction associated with the magnetic transition 
is easily observed.

Taken together, the x-ray results in Fig.~2(c) and the neutron 
diffraction extinction release profile in Fig.~2(d) allow us to 
refine the assignment of the structural transition in resistivity 
($R$) measurements.  In previous work\cite{Pratt09,Ni09}, $T_S$ 
was estimated from the first deviation in d$R$/d$T$ from its 
high-temperature behavior.  Here, we see that $T_S$ is most 
appropriately associated with the first minimum in d$R$/d$T$ 
which falls at a slightly lower temperature.

\begin{figure}
\includegraphics[width=1.0\linewidth]{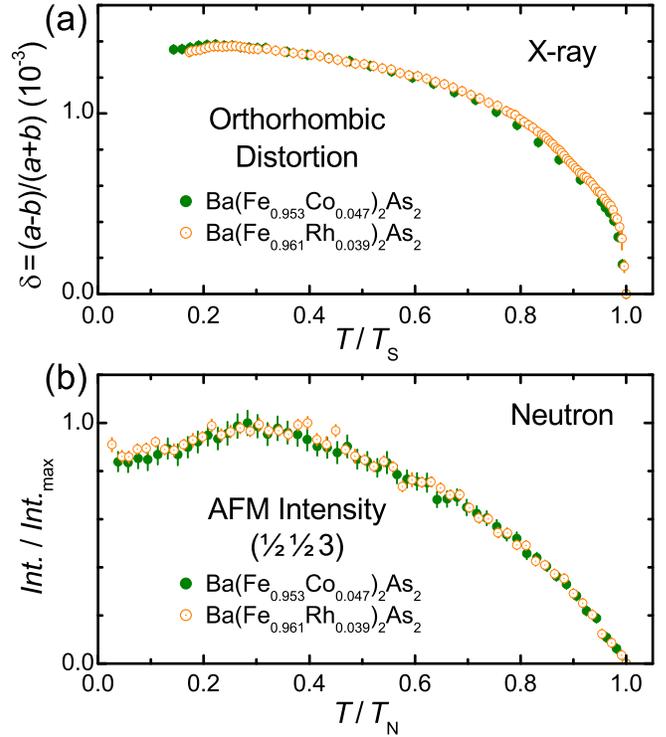}
\caption{\label{fig:fig4}(a) Measured orthorhombic distortions 
for Ba(Fe$_{0.961}$Rh$_{0.039}$)$_2$As$_2$ and 
Ba(Fe$_{0.953}$Co$_{0.047}$)$_2$As$_2$ as function of reduced 
temperature $T$/$T_S$. (b) Integrated intensity of the 
(1/2\,1/2\,3) magnetic reflections for 
Ba(Fe$_{0.961}$Rh$_{0.039}$)$_2$As$_2$ and 
Ba(Fe$_{0.953}$Co$_{0.047}$)$_2$As$_2$ normalized to their 
maximum value at $T_c$ plotted as function of reduced temperature 
$T$/$T_N$.}
\end{figure}

Magnetic reflections were observed at 
\textbf{Q$_{\textrm{AFM}}$}~= (1/2\,1/2\,$L$) positions with $L$ 
odd.  Scans at several representative temperatures are shown in 
Fig.~3(b).  From the data at 57\,K we see that the AFM order sets 
in at lower temperatures than the orthorhombic distortion 
illustrated by the finite splitting of the (1\,1\,10) reflection 
in Fig.~3(a) without the appearance of a magnetic (1/2\,1/2\,$L$) 
reflection in Fig.~3(b).  The magnetic propagation vector is 
identical to that for BaFe$_2$As$_2$ and the 
Ba(Fe$_{1-x}$Co$_x$)$_2$As$_2$ compounds, indicating that the 
magnetic structure is the same AFM ``stripe'' structure observed 
for all of the AFM ordered iron arsenides.  Analysis of the 
intensity ratios of different AFM reflections at selected 
temperatures confirms that the moment direction is the same, 
along the elongated orthorhombic \textbf{a} direction.

Figure~2(d) shows the magnetic order parameter obtained from the 
intensity measured at \textbf{Q$_{\textrm{AFM}}$}~= (1/2\,1/2\,3) 
as function of temperature for 
Ba(Fe$_{0.961}$Rh$_{0.039}$)$_2$As$_2$.  Above $T_N$~= 54\,K, no 
scattering is observed at \textbf{Q$_{\textrm{AFM}}$}, but 
increases smoothly below $T_N$.   Furthermore, lowering the 
temperature below $T_c$~= 14\,K results in a decrease in the 
magnetic order parameter in the same manner, and nearly the same 
magnitude, as observed for 
Ba(Fe$_{0.953}$Co$_{0.047}$)$_2$As$_2$, indicating that the 
strong coupling between superconductivity and magnetism is not 
exclusive to Ba(Fe$_{1-x}$Co$_x$)$_2$As$_2$ compounds.  Using the 
measured mass of the sample and assuming that the entire sample 
contributes to the scattering at \textbf{Q$_{\textrm{AFM}}$}~= 
(1/2\,1/2\,3), the magnetic scattering data were normalized using 
the parent BaFe$_2$As$_2$ compound as a standard through the 
process described in Refs.\cite{Pratt09,Fernandes10}.  The 
extrapolated ordered moment  at zero-temperature was determined 
to be (0.37$\pm$0.1)\,$\mu_B$, somewhat larger than the value 
(0.25$\pm$0.1)\,$\mu_B$ determined for 
Ba(Fe$_{0.953}$Co$_{0.047}$)$_2$As$_2$\cite{Pratt09,Fernandes10}, 
but well below the value (0.87$\pm$0.1)\,$\mu_B$ for the parent 
BaFe$_2$As$_2$ compound\cite{Huang08}. 

Taken together, these data clearly show that the suppression of 
the magnetic and structural order parameters previously observed 
in Ba(Fe$_{1-x}$Co$_x$)$_2$As$_2$ in the superconducting regime 
are also observed in Ba(Fe$_{0.961}$Rh$_{0.039}$)$_2$As$_2$. This 
point is most clearly made in Fig.~4, which plots the magnitude 
of the orthorhombic distortion [Fig.~4(a)] and the intensity of 
the (1/2\,1/2\,3) magnetic diffraction peaks [Fig.~4(b)] for both 
Ba(Fe$_{0.953}$Co$_{0.047}$)$_2$As$_2$ and 
Ba(Fe$_{0.961}$Rh$_{0.039}$)$_2$As$_2$.  Here we have plotted the 
data in terms of the reduced temperature scale $T$/$T_S$ and 
$T$/$T_N$, respectively, to emphasize the close similarity in the 
behavior of these compounds manifested in: (i) similar 
temperature dependence for the magnetic and structural order 
parameters; (ii) the similar absolute value of the distortion 
and; (iii) the similar qualitative and quantitative effects of 
superconductivity on both order parameters.  These results 
suggest that the strong interaction between magnetism, structure 
and superconductivity is a general feature of electron-doped 
BaFe$_2$As$_2$ superconductors.\\

During preparation of this manuscript, a neutron scattering study 
in Ba(Fe$_{1-x}$Ni$_x$)$_2$As$_2$ compounds has appeared that 
shows a suppression of the AFM signal below $T_c$ for an 
underdoped sample confirming our previous 
results\cite{Pratt09,Fernandes10} and the present 
work.\cite{Wang10}\\

We thank A.~Kracher for performing the WDS measurements.  This 
work was supported by the U.\,S. Department of Energy, Office of 
Basic Energy Science, Division of Materials Sciences and 
Engineering. The research was performed at the Ames Laboratory. 
Ames Laboratory is operated for the U.\,S. Department of Energy 
by Iowa State University under Contract No. DE-AC02-07CH11358.  
Work at Oak Ridge National Laboratory is supported by U.\,S. 
Department of Energy, Office of Basic Energy Sciences, Scientific 
User Facilities Division.


\begin{thebibliography}{}
\bibitem[()]{Huang08}Q.~Huang, Y.~Qiu, W.~Bao, J.\,W.~Lynn, M.\,A.~Green, Y.~Chen, T.~Wu, G.~Wu, and X.\,H.~Chen, Phys. Rev. Lett. \textbf{101}, 257003 (2008).
\bibitem[()]{Jesche08}A.~Jesche, N.~Caroca-Canales, H.~Rosner, H.~Borrmann, A.~Ormeci, D.~Kasinathan, H.\,H.~Klauss, H.~Luetkens, R.~Khasanov, A.~Amato, A.~Hoser, K.~Kaneko, C.~Krellner, and C.~Geibel, Phys. Rev. B \textbf{78}, 180504(R) (2008).
\bibitem[()]{Goldman08}A.\,I.~Goldman, D.\,N.~Argyriou, B.~Ouladdiaf, T.~Chatterji, A.~Kreyssig, S.~Nandi, N.~Ni, S.\,L.~Bud'ko, P.\,C.~Canfield, and R.\,J.~McQueeney, Phys. Rev. B \textbf{78}, 100506(R) (2008).
\bibitem[()]{Ni08}N.~Ni, S.~Nandi, A.~Kreyssig, A.\,I.~Goldman, E.\,D.~Mun, S.\,L.~Bud'ko, and P.\,C.~Canfield, Phys. Rev. B \textbf{78}, 014523 (2008).
\bibitem[()]{Sefat08}A.\,S.~Sefat, R.~Jin, M.\,A.~McGuire, B.\,C.~Sales, D.\,J.~Singh, and D.~Mandrus, Phys. Rev. Lett. \textbf{101}, 117004 (2008).
\bibitem[()]{Ni08b}N.~Ni, M.\,E.~Tillman, J.-Q.~Yan, A.~Kracher, S.\,T.~Hannahs, S.\,L.~Bud'ko, and P.\,C.~Canfield, Phys. Rev. B \textbf{78}, 214515 (2008).
\bibitem[()]{Chu09}J.-H.~Chu, J.\,G.~Analytis, C.~Kucharczyk, and I.\,R.~Fisher, Phys. Rev. B \textbf{79}, 014506 (2009).
\bibitem[()]{Lester09}C.~Lester, J.-H.~Chu, J.\,G.~Analytis, S.~Capelli, A.\,S.~Erickson, C.\,L.~Condron, M.\,F.~Toney, I.\,R.~Fisher, and S.\,M.~Hayden, Phys. Rev. B \textbf{79}, 144523 (2009).
\bibitem[()]{Pratt09}D.\,K.~Pratt, W.~Tian, A.~Kreyssig, J.\,L.~Zarestky, S.~Nandi, N.~Ni, S.\,L.~Bud'ko, P.\,C.~Canfield, A.\,I.~Goldman, and R.\,J.~McQueeney, Phys. Rev. Lett. \textbf{103}, 087001 (2009).
\bibitem[()]{Christianson09}A.\,D.~Christianson, M.\,D.~Lumsden, S.\,E.~Nagler, G.\,J.~MacDougall, M.\,A.~McGuire, A.\,S.~Sefat, R.~Jin, B.\,C.~Sales, and D.~Mandrus, Phys. Rev. Lett. \textbf{103}, 087002 (2009).
\bibitem[()]{Fernandes10}R.\,M.~Fernandes, D.\,K.~Pratt, W.~Tian, J.~Zarestky, A.~Kreyssig, S.~Nandi, M.\,G.~Kim, A.~Thaler, N.~Ni, P.\,C.~Canfield, R.\,J.~McQueeney, J.~Schmalian, and A.\,I.~Goldman, arXiv:0911.5183 (unpublished).
\bibitem[()]{Nandi10}S.~Nandi, M.\,G.~Kim, A.~Kreyssig, R.\,M.~Fernandes, D.\,K.~Pratt, A.~Thaler, N.~Ni, S.\,L.~Bud'ko, P.\,C.~Canfield, J.~Schmalian, R.\,J.~McQueeney, and A.\,I.~Goldman, Phys. Rev. Lett. \textbf{104}, 057006 (2010).
\bibitem[()]{Fang08}C.~Fang, H.~Yao, W.-F.~Tsai, J.~Hu, and S.\,A.~Kivelson, Phys. Rev. B \textbf{77}, 224509 (2008).
\bibitem[()]{Xu08}C.~Xu, M.~M\"uller, and S.~Sachdev, Phys. Rev. B \textbf{78}, 020501(R) (2008).
\bibitem[()]{Fernandes10b}R.\,M.~Fernandes, L.\,H.~VanBebber, S.~Bhattacharaya, P.~Chandra, V.~Keppens, D.~Mandrus, M.\,A.~McGuire, B.\,C.~Sales, A.\,S.~Sefat, and J.~Schmalian, arXiv:0911.3084 (unpublished).
\bibitem[()]{Canfield10}P.\,C.~Canfield and S.\,L.~Bud'ko, arXiv:1002.0858 (unpublished).
\bibitem[()]{Canfield09}P.\,C.~Canfield, S.\,L.~Bud'ko, N.~Ni, J.\,Q.~Yan, and A.~Kracher, Phys. Rev. B \textbf{80}, 060501 (2009). 
\bibitem[()]{Ni09}N.~Ni, A.~Thaler, A.~Kracher, J.\,Q.~Yan, S.\,L.~Bud'ko, and P.\,C.~Canfield, Phys. Rev. B \textbf{80}, 024511 (2009). 
\bibitem[()]{Wang10}M.~Wang, H.~Luo, J.~Zhao, C.~Zhang, M.~Wang, K.~Marty, S.~Chi, J.\,W.~Lynn, A.~Schneidewind, S.~Li, P.~Dai, arXiv:1002.3133 (unpublished). 
\end{thebibliography}
\end{document}